\documentclass[useAMS,usenatbib]{mn2e}
\pdfoutput=1 

\usepackage{graphicx}
\usepackage{amsmath,amssymb}
\usepackage{bm}
\usepackage{url}

\newcommand{\imgwidth}{3.5in} 
\newcommand{\F}{{\mathcal{F}}}
\newcommand{\G}{{\mathcal{G}}}
\newcommand{\fmap}{\mathcal{F}_{\text{aper}}}
\newcommand{\Sersic}{S{\'e}rsic}

\begin{document}
\title{Flexion in Abell 2744}
\date{}
\author[Bird, Goldberg]
{J. P.~Bird,$^1$\thanks{justin.bird@drexel.edu}
D. M.~Goldberg$^1$
\\
$^1$Department of Physics, Drexel University\\
Philadelphia, PA 19104\\
}

\maketitle


\begin{abstract}

We present the first flexion-focused gravitational lensing analysis of 
the first of the strong-lensing ``cosmic telescope" galaxy clusters, 
observed as part of the \emph{Hubble Frontier Fields} initiative. 
Using HST observations of Abell 2744 (z = 0.308), 
we apply a modified Analytic Image Model (AIM) 
technique to measure source galaxy flexion and shear values at a 
final number density of 82 arcmin\(^{-2}\).
By using flexion data alone we are able to identify the primary 
mass structure aligned along the heart of the cluster 
in addition to a major substructure peak offset \(1.43'\) from the 
cluster core. 
We generate two types of nonparametric reconstructions: 
a flexion aperture mass map, which identified the 
central potential and substructure peak with mass signal-to-noise 
of \(3.5\sigma\) and \(2.3\sigma\) respectively; and a convergence 
map derived directly from the smoothed flexion field.
For the primary peak we find a mass of 
\(1.93\times10^{14}\,h^{-1}\,M_{\odot}\) within a 45" (145\(h^{-1}\) kpc)
aperture, and for the western substructure we find a mass of 
\(7.12\times10^{13}\,h^{-1}\,M_{\odot}\) within a 
25" (80\(h^{-1}\) kpc) aperture. The associated peak velocity
dispersions were determined to be \(\sigma_v = 1630\) km/s 
and \(\sigma_v = 766\) km/s, respectively, by fitting nonsingular
isothermal sphere profiles to the flexion data.
Additionally, we use simultaneous shear measurements 
to independently reconstruct the broader cluster mass structure, and 
find that it is unable to reproduce the small-scale structure associated with 
the flexion reconstructions.
Finally, we perform the same analysis on the Abell 2744 parallel sky field, 
and find no strong phantom signals in the noise reconstructions.

\end{abstract}



\section{Introduction}

Over the last two decades, gravitational lensing has served as one of the most successful tools in the development of our standard model of cosmology and extragalactic astronomy \citep[e.g.][etc.]{Blandford1992,KS1993,KSB1995,Klypin/Kravtsov,Wittman2000,VanWaerbeke2000,Bartelmann/Schneider,Kravtsov2004,
Lewis2006,Mandelbaum2006,Munshi2008,Tinker2008,Massey2010,Kneib2011,Hoekstra2013} 
Galaxy clusters are ideal candidates for observing gravitational lensing effects in background sources, as their high mass density and broad angular extent ensure a large population of lensed images. 
By comparing the differential distortion patterns of lensed sources across the extent of the cluster, one is able to produce a mapping of the smoothed cluster mass density distribution on arcminute scales \citep{Kneib1996,Bacon2000,CloweBullet}.
Indeed gravitational lensing has already been widely applied to map the extended mass distribution of many other galaxy clusters in this manner, such as in \cite{Deb2010,Evans2009,Okabe2010}.  

While most weak lensing reconstructions focus on questions of total mass, radial profiles, and the measurement of cluster ellipticities, it is well-known from both numerical simulations \citep[e.g.][]{Klypin/Kravtsov,Diemand} and semi-analytical work \citep{Press/Schechter,Sheth/Tormen99} that significant substructure is expected in dark matter halos at approximately the 2\% level. 
Correspondingly, the advent of larger and deeper datasets now allows precise and accurate mass reconstructions in both the inner core and outer regions of galaxy clusters, through strong lensing and combined approaches \citep[e.g.][]{Bradac2004,Leonard2007,Jauzac2015_2}. 
Strong lensing techniques are able to probe the extremely dense cluster core by position matching multiply-imaged systems, and have achieved galaxy-scale mass resolution with both parametric (mass-follows-light assumptions) and non-parametric reconstructions. 
In particular, the last few years have seen an explosion of new strong (+weak) gravitational lensing analyses, particularly through the data obtained in the recent Hubble Frontier Fields\footnote{\url{http://www.stsci.edu/hst/campaigns/frontier-fields/}} (HFF) initiative
\citep[e.g.][etc.]{Jauzac2014,Jauzac2015_1,Jauzac2015_2,Jauzac2016,Mo2016,
Diego2015_1,Diego2015_2,Wang2015,Zitrin2013,Zitrin2015,Sharon2015,
Grillo2015,Lam2014,Johnson2014,Richard2014,Sendra2014,Medezinski2013}.

While strong lensing techniques can probe the extremely dense core of clusters, in general it is very difficult to detect substructure in the \((10^{12}\)-- \(10^{13})\,M_{\odot}\) regime. 
We propose that one of the best ways to uncover the detailed structure of dynamic clusters is to look at the higher-order lensing statistic known as ``flexion" \citep{Goldberg/Natarajan,Goldberg2005,Bacon2006}. 
Flexion induces arcing effects in lensed galaxies, similar to but smaller in scale than the giant arcs of strong lensing, forming ``arclets" near local mass overdensities. 
The advantage to using this higher-order information is that it is both far more sensitive to small-scale perturbations of the convergence field than shear, and is viable further away than the radial distance within which strongly-lensed giant arcs would be produced. 
Jointly, the lack of a detectable flexion signal in a region can impose constraints on the shape and size of any possible local substructure. 

With strong- and weak- lensing groups already taking advantage of the unprecedented density and depth of the HFF, there is enormous potential for applying flexion analyses to this new generation of observations. Flexion as an indicator of local structure will provide significantly more information if there is a high density of sources, and thus the HFFs are in a unique position to calibrate flexion measurement and reconstruction techniques for other, wider surveys and applications.

In this paper we present a gravitational lensing flexion analysis of the widely-studied galaxy cluster Abell 2744, detailing our methodology from image data, to measured flexion/shear, to final reconstruction maps of the cluster convergence field and estimates of structure mass. We show the efficacy of using only flexion in determining the large and intermediate scale structure of the cluster, and compare our flexion-convergence reconstruction with a shear-convergence reconstruction calculated using the same dataset. Additionally, we produce flexion aperture mass maps and investigate similarities between the two types of reconstructions.

In order to determine any systematic effects in our measurement procedure, we perform a simultaneous flexion analysis of the HFF Abell 2744 parallel field, chosen specifically to contain no significant mass structures with HST observations taken under similar conditions, and we compare the resultant noise reconstructions with the cluster field.

Section \S\ref{sec:background} provides a background of the lensing formalism used in this work, and the dataset, observational pipeline, and data reduction process. In Section \S\ref{sec:methods} we summarize our analysis methodology, including details and implementation of flexion measurement and our mass estimators. Section \S\ref{sec:results} presents the results our analysis on Abell 2744 and its associated parallel field including derived convergence and mass signal maps, concluding with a summary and discussion.

We adopt the standard conventions of \(\Lambda\)CDM cosmology with \(\Omega_m = 0.3, \Omega_{\Lambda} = 0.7\), Hubble constant \(H_0 = 100h \) km/s/Mpc, and unless otherwise specified, all magnitudes are listed in the AB system. Likewise, we adopt complex notation for denoting directional vectors and define the gradient operator as \(\partial = \partial_1 + i\partial_2\). 


\section{Background}
\label{sec:background}

\subsection{Lensing and flexion formalism}

Gravitational flexion was originally developed as a way to quantify gradients in the dimensionless surface mass density field, \(\kappa\),
through highly distorted, gravitationally lensed ``arclet" images \citep{Goldberg/Natarajan,Goldberg2005,Bacon2006}. This characteristic bending and arcing in a background source's image has been described as ``banananess" \citep{Schneider/Er}. Flexion is particularly powerful in that it probes the third derivative of the underlying potential, and thus, it is far more sensitive than shear to substructure in cluster halos while not requiring the multiple images necessary for a full strong lensing analysis.

To reconstruct the lens-plane convergence, \(\kappa\), with measurable galaxy-image properties, we can establish a normalized foreground mass potential \(\psi\) through the Poisson equation:
\begin{equation}\label{eqn:potential}
\partial^*\partial \psi = 2\kappa
\end{equation}
with derivatives in angular units in the lens (galaxy cluster) plane. This potential may now be related to a number of observables, including the complex shear field,
\begin{equation}
\gamma = \gamma_1 + i\gamma_2 = \frac{1}{2}\partial\partial\psi,
\end{equation}
which induces tangential elongation to a source image. 
Spin-1 flexion, \(\F\), has a natural interpretation as the gradient of the convergence, inducing distortion that ``skews" source images (coma), while spin-3 flexion, \(\G\), has \(m = 3\) rotational symmetry and acts as the gradient of the shear, producing ``arcing" effects (trefoil) in an image (see fig. 1 \citet{Bacon2006}):
\begin{align}
\F &= |\F|e^{i\phi} = \frac{1}{2}\partial\partial^*\partial\psi = \partial \kappa,\label{eqn:F}\\
\G &= |\G|e^{i3\phi} = \frac{1}{2}\partial\partial\partial\psi = \partial \gamma .\label{eqn:G}
\end{align}
In the weak lensing regime (\(\kappa \ll 1\)), a lensing distortion can be represented by the linear combination of these and potentially higher-order effects. In practice however, detecting flexion and higher-order distortions in typical lensing configurations requires an instrument with both a small point spread function (PSF) and a very high angular resolution, failing which the precise shape information is lost in pixel noise. In this regard, one of the most effective ways to observe these lensing effects is to avoid an atmospheric PSF in the first place through space-based observations. In particular, the recent Hubble Space Telescope Frontier Fields project has produced the deepest observations of lensing clusters yet attained, and the second-deepest observations of blank fields, offering one of the best opportunities for flexion study yet.


\subsection{Observations and Data}
\label{sec:data}

With a total of 70 HST orbits for each of six massive cluster lenses and 
each respective parallel field, the goal of the Hubble Frontier Field program 
is to probe the early universe at redshifts up to and including the \(z = (6\)--\(8)\) 
regime, with a limiting magnitude of mag\(_{AB} \sim 28.7-29\) in seven 
passbands for a \(5\sigma\) point source detection in a 0.4" aperture. At this redshift, assuming
the previously stated cosmology gives a scale of \(3.175h^{-1}\) kpc/arcsec 
(or 4.536 kpc/arcsec if assuming \(h = 0.7\)).

The angular scale of the data (0.05" pixel scale, 0.03" after drizzling), 
the abundance of color information, the considerable mass 
(\citet{Jauzac2015_1} measures the core to have a mass of 
\(2.76 \times 10^{14} M_{\odot}\) within 250 kpc), and favorable 
orientations of the cluster lenses create favorable conditions for 
strong flexion signal. The first cluster to be fully imaged was Abell 2744 
at a redshift of z=0.308 with a field size of 3.5' x 3.5' \citep{Lotz2016}, 
and it is to this first cluster that we apply our flexion pipeline. 
Taking advantage of the high-level science images
\footnote{\url{https://archive.stsci.edu/pub/hlsp/frontier/abell2744/images/hst/}} 
released to the MAST archives from the HST Frontier Fields Science 
Products Team, we use the WFC3 ACS F435w, F616w, and F814w 
filters for color magnitudes, and make shape measurements in the 
F814w filter, with total respective filter integration times corresponding 
to 24, 14, and 46 orbits, respectively.


\section{Methods}
\label{sec:methods}

\subsection{Data Reduction}
\label{sec:data_reduction}

Flexion analyses pose unique challenges as compared to weak lensing shear. If we consider the simplified case of a singular isothermal sphere (SIS), the shear and flexion signal strength fall off as \(1/\theta\) and \(1/\theta^2\), respectively. While a shear reconstruction may require averaging over hundreds of galaxies to produce a significant signal, a flexion reconstruction might only utilize a handful, and thus requires a much higher S/N ratio per galaxy.

To optimize the quality of our galaxy sources and the accuracy of their 
associated uncertainty-maps in preparation for flexion measurement, 
further handling and manipulation of the STScI-calibrated HST 
images is necessary. The process of identifying source galaxies down 
to the background level, deblending a dense cluster field, extracting individual 
galaxies while masking other contaminating sources, and targeting the
correct sources for measurement is nontrivial. 
We create a post-processing pipeline to optimize the specifics of our data 
creation, combining it with the \texttt{Source Extractor} \citep{SExtractor} 
utility which can be used for identifying background signatures that match 
specified criteria, namely involving pixel thresholding levels, aperture 
size, deblending, and noise parameters.

Alongside the identification of galaxy objects in the field, there are selection cuts that can be made both before and after measurements are made to decrease possible contaminants in the flexion signal, and which depend on the inherent and apparent properties of each source. For example, flexion signal depends on the gradient of the lens convergence and is therefore very sensitive to the apparent size of the lensed source, causing larger objects to be preferentially selected. Additionally, the intrinsic shape of a galaxy can play a significant role in a measurement of its ``intrinsic flexion" -- the flexion that would be measured if the object was not gravitationally lensed at all. Irregular galaxies in particular can mimic a strong external flexion signal. As the flexion signal can depend on far fewer objects than shear, data reduction and source selection is an extremely important component of this analysis. 

Our data reduction strategy involves the following general steps:
\begin{enumerate}
\item We crop all images to exclude regions which have poor stacking coverage, decreasing background estimation bias in \texttt{Source Extractor} and increasing performance.\\
\item Using redshift and magnitude catalogs \citep{Owers2011,Merten2011}, we position-match and identify any established objects belonging to either the cluster itself or the foreground. We identify the foreground-associated pixels down to background level and replace them with a randomized noise map set to the mean of the local background. \label{step:pixel_replace}\\
\item We implement a ``hot-cold" strategy \citep{Rix2004,Leonard2007} running \texttt{Source Extractor} in dual-image mode across all appropriate bands - first clearing the field of large bright objects not previously removed, then targeting the expected source population with a minimum of 15 pixels over a \(2\sigma\) detection threshold. \\
\item As spectroscopic and photometric redshifts are available for only a fraction of our detected objects, we also use a series of criteria to exclude probable source contaminants or low S/N objects. Namely, we reject bright or large galaxies with  \(mag_{\,\mathrm{F814w}} < 24\) or \(\mathrm{FWHM} > 0.9"\), establish a low signal bound by excluding galaxies with \(mag_{\,\mathrm{F814w}} > 28\), \(\mathrm{FWHM} < (2\times\mathrm{FWHM}_{\,\mathrm{PSF}})\), or S/N ratio \(< 20\). Additionally, we remove objects which are flagged as incomplete (FLAG \(\ge8\)) in the \texttt{Source Extractor} catalog, indicating closely associated or fragmented objects, and mark objects which have FLAG totals \(<8\) to use as a discriminant in later visual inspection (see section \ref{sec:stampcheck}).\\
\item We generate square postage stamps of selected sources. These are 
centered at the galaxy centroids and have a windowed radial extent set to 
4.5 times the calculated half-light radii, chosen to be large enough to 
ensure that the flexion-susceptible galaxy wings and background 
zero-constraints are included. \\
\item Alongside the image stamp we generate an associated \(1\sigma\) 
noise stamp which includes both background sky noise and Poisson noise. 
\end{enumerate}

\subsection{Flexion Measurement - Analytic Image Modeling (AIM)}
\label{sec:measurement}

While research into the potential power of flexion measurement continues to remain popular \citep[][etc.]{Flexion/Viola,Er2012,Cain2016,Cardone2016}, there have only been a handful of flexion analyses applied to real data to date \citep[]{Leonard2007,Leonard2011,Okura2008,Cain2011,Cain2015}, with several using the same dataset and the majority in the widely studied rich cluster Abell 1689. 

Flexion measurements of real, individual sources have been achieved through a few techniques, including by measuring a combination of various third (and higher) order moments of the light distribution (as originally suggested by \citet{Goldberg/Natarajan,Okura2008}), decomposing the projected galaxy shape onto a polar orthonormal ``shapelets'' basis set and truncating the series at a particular threshold \citep{Massey2007,Goldberg/Leonard}, and exploring the local potential field through parameterized ray-tracing \citep{Cain2011}, known as Analytic Image Modeling (AIM).

We utilize a modified version of the AIM technique to determine the localized flexion field values for individual sources. AIM is distinct in that instead of measuring derived quantities (such as weighted surface brightness moments), it fits the lensed galaxy objects using a parametric model. By comparing the observed image to the uncertainty-weighted model image, we are able to optimize the parameters over reasonable bounds and thus constrain the flexion fields (and other shape information.)

\subsubsection{Model Parameterization and Suitability}

Galaxy profiles are largely fit to parametric models with a radial luminosity 
distribution, and for this work we implement an elliptical \Sersic{} intensity profile. 
The \Sersic{} profile is particularly useful in that it encompasses a range of 
different models, including ones already well-established through 
galaxy-luminosity profiling, including exponential, Gaussian, and 
de Vaucoulers profiles. Our model takes the form:
\begin{equation}
I(\theta) = I_e \exp{\left\{-b_n\left[\left(\frac{\theta}{\theta_e}\right)^{1/n}-1\right]\right\}},
\end{equation}
where \(\theta_e\) is the radius of the isophote containing half the total 
flux of the galaxy (the half-light radius), \(I_e\) is the brightness at this 
effective radius, and \(n\) is the S{\'e}rsic index, which controls the 
steepness of the profile. 
We use the total integrated surface brightness with analytic form:
\begin{equation}
L_{\mathrm{tot}} = 2\pi n I_e \theta_e^2 \frac{e^{b_n}}{(b_n)^2}\,\Gamma(2n) \,q
\end{equation}
as the optimized flux parameter to ensure that both the fit and source 
galaxy are contained within stamp boundaries. 
The profile's radial behavior is described by
\begin{equation}
\theta = \sqrt{\left(\frac{(\theta_1-\theta_{1,0})'}{q}\right)^2 + \left((\theta_2-\theta_{2,0})'\right)^2}
\end{equation}
where the primes indicate a frame rotated by angle \(\phi\), and \(q\) is the galaxy's principal axis ratio. 
We decompose the axis ratio and rotation angle using the following definition of the (third) eccentricity:
\begin{equation}
|e| \,=\, \frac{q^2 - 1}{q^2 + 1}
\end{equation}
Finally, \(b_n\) is empirically determined to set \(\theta_e\) to the half-light isophotal radius, 
and we adopt an approximated functional form
\begin{equation}
b_n = 1.992n - 0.3271,\quad 0.5 < n < 8.0
\end{equation}
valid within the given \Sersic{} index range \citep{Capaccioli1989}.

To circumvent the shear/ellipticity degeneracy we fix the lensing 
shear \(\gamma\) to 0.0 and allow the intrinsic ellipticity to absorb 
the two degenerate parameters into one. Although it is possible to 
set the lensing shear to that of a predetermined model, instead we 
aim to obtain an unbiased estimate of the shear field alongside the 
targeted flexion fields. 

Overall, there are seven model parameters for the unlensed galaxy 
light profile, and four effective parameters for the lensing fields:
\begin{equation}
\{n,L_{tot},q,\phi,\theta_e,\theta_{0,1},\theta_{0,2}, \F_1,\F_2,\G_1,\G_2\}.
\end{equation}
While we can reasonably expect this model to work well for simple source
galaxies, it is also important to explore typical parameter spaces for existing
data and to consider any known limitations.

\subsubsection{Typical Model Suitability}

One of the biggest problems for flexion measurement in general is the inherent inhomogeneity of galaxy structure, particularly when combined with the emphasis on a select few sources (again as a result of strong local mass sensitivity). A complete analysis of ``inherent flexion" in unlensed populations has yet to be fully investigated at this point, though previous analyses have done so in cluster lensed populations \citep{Leonard2007,Okura2008,Cain2011}. Most galaxies are not simply circular or elliptical, and many can be expected to contain structural irregularities such as arms, bars, or star-forming regions. Elliptical models tend to have a large range of \Sersic{} indices, whose values depend strongly on luminosity \citep{Blanton2009}. Furthermore, ellipticals can also be separated into ``boxy" and ``disky" subtypes. 

For the most part, however, a \Sersic{} model can be used to model the overall structure of most galaxy types. For elliptical galaxies, the \Sersic{} index \(n\)  generally reflects what is apparently a single component galaxy. Spiral galaxies can also be described by a \Sersic{} index, but in this case \(n\) reflects a balance between the disk and the bulge, two clearly distinct components. Morphology typically focuses on the separation of the disk from the bulge, usually treating the disk as an exponential profile and fitting the bulge to a \Sersic{} profile, with \(n\) dependent on the type of the central component. Along with lenticular galaxies, these form the basis of our unlensed galaxy reconstruction efforts. The range of possible parameters create a broad distribution of shapes; drawing a good statistical likelihood of inherent shape is difficult. In this work we aim to partially investigate inherent flexion by performing a simultaneous flexion analyses on the offset parallel Abell 2744 field.

\subsubsection{PSF Correction}
\label{sec:psf}

Because of the forward-modeling nature of the AIM technique, a PSF correction can be readily applied during parameter minimization. Our images consist of many offset, rotated, and stacked individual exposures over noncontiguous time periods (with a total average integration time of over 40,000 seconds, or \(\sim11\) hours, in the  band), increasing the object S/N by a factor roughly equivalent to the number of orbits (46 in F814w) at the cost of a more complicated effective PSF which is nontrivial to model. The 90-minute HST orbital period introduces thermal breathing fluctuations while the telescope focus deviates over periods of weeks, and our data spans a range of these cycles. Furthermore, the off-axis position of the ACS introduces a spatially-varying geometric distortion across the WFC chips.

\begin{figure}
\begin{center}
\includegraphics[width=3.3in]{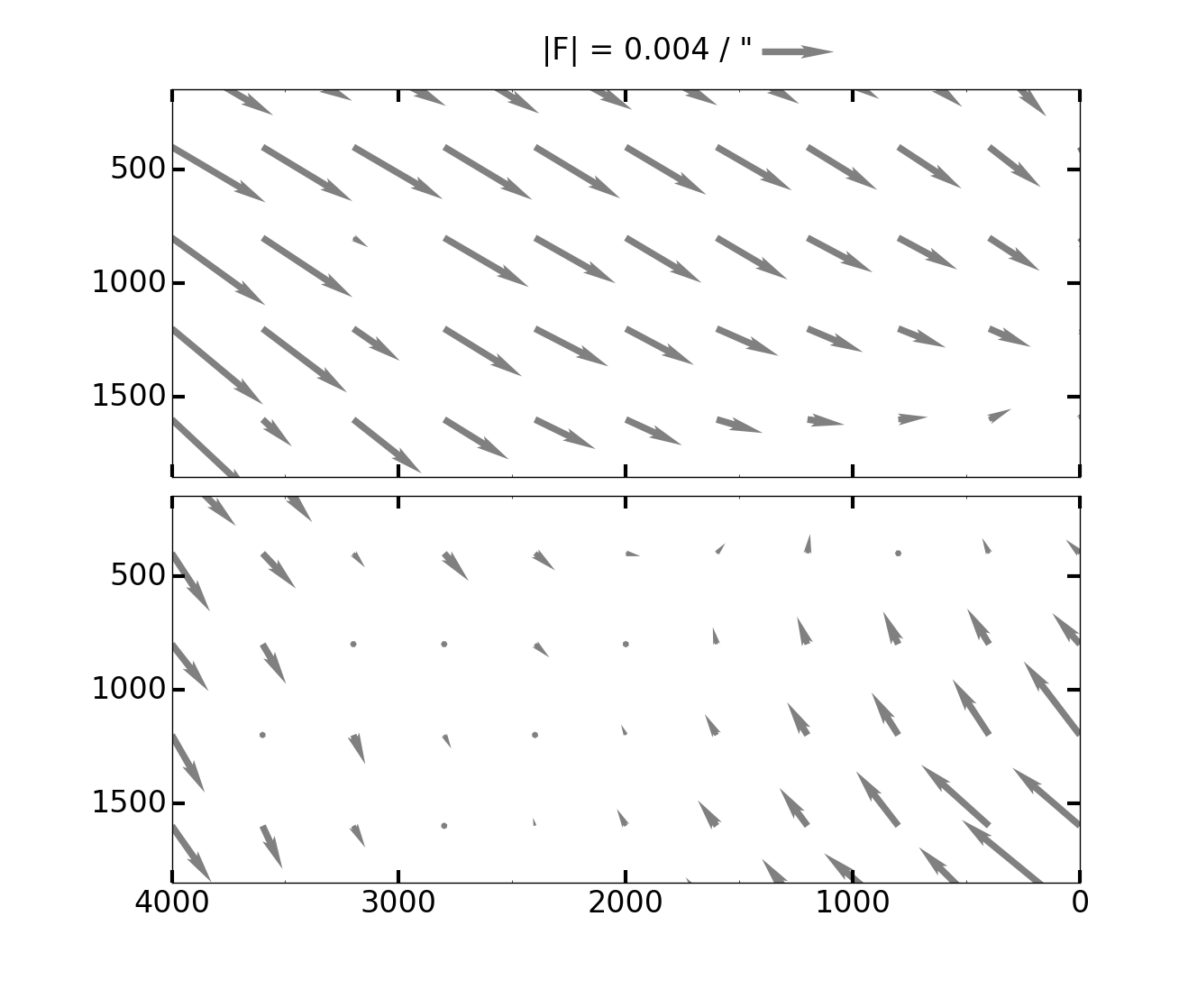}
\end{center}
\caption[]{Measured \(\F\)-flexion vectors of the simulated 
ACS WFC PSF across chips 2 and 1. 
We use TinyTim to model each PSF, 
vary the input spatial coordinates to create a grid 
representing a single exposure, and measure the flexion
signal using our AIM implementation. In practice, stacking the slightly offset 
and rotated multiple exposures reprojects 
the images and reduces the directional bias of the PSF.}
\label{fig:psf_flexion}
\end{figure}

While there are techniques that have been successfully applied to model the PSF of stacked ACS images \citep[e.g.][]{Bacon2003,Rhodes2007,Harvey2015,Jauzac2016}, these have mostly found use in shear measurement, with accuracy requirements at the \(\sim\)1\% level. 
As the induced flexion in a source from the PSF is 
\begin{equation}\label{eqn:psf}
\F_{\mathrm{induced}} \sim \F_{\mathrm{PSF}} \frac{a_{\mathrm{PSF}}^4}{a_{\mathrm{source}}^4 + a_{\mathrm{PSF}}^4}
\end{equation}
as derived in \citet{Leonard2007}, for typical WFC values 
(\(\F_{\mathrm{PSF}} \sim 10^{-3}/",  a_{\mathrm{PSF}} \sim 0.1125"\)) 
there is not a significant flexion contribution from the PSF, 
provided the source is sufficiently large. 
As one of our pre-measurement cuts requires \texttt{Source Extractor} sources to have a FWHM greater than twice that of the ACS PSF (\(2\times 0.1125"\)), this minimizes the amount of any possible PSF-induced flexion.
To test the flexion anisotropy across the ACS WFC chips, we used the 
TinyTim\footnote{\url{http://www.stsci.edu/hst/observatory/focus/TinyTim}}
software to simulate an ACS PSF under typical 
instrument parameters. By varying the input pixel coordinates across the
two chips, we created a grid of spatially-varying PSFs representing a single
exposure, and were then able to measure the flexion signal using our AIM 
implementation. Figure \ref{fig:psf_flexion} shows the resultant \(\F\)-flexion 
vectors, which have a maximum magnitude of \(|\F| = 0.005\), much smaller than 
the lensing flexion signal expected in Abell 2744, particularly in sources much
larger than the PSF. In addition, stacking the offset and rotated individual exposures 
into the integrated image that we use reprojects and averages the directional biases 
of the flexion anisotropies.

In practice, as an approximation we use a simple Gaussian convolution with FWHM \(= 2\sqrt{2\mathrm{ln}\,2}\sigma = 0.1125"\) to account for the general smoothing of a PSF. We expect the PSF to have relatively small effects on the overall shape and shape parameters of our selected galaxies, but as a self-consistency check we perform the best-fit again without any applied PSF. If the fit shape or model parameters are markedly different we exclude that object from our analyses.

\subsubsection{Implementation}

To determine the best-fit model parameterization we use a 
Levenberg-Marquardt least-squares minimization scheme, 
the core of which is implemented through a modified version 
of the Python 
translation\footnote{\url{https://code.google.com/archive/p/astrolibpy/}} 
of the IDL code \texttt{mpfit}. Our pipeline combines this minimization 
technique with galaxy models simulated through the 
\texttt{GalFlex}\footnote{\url{http://physics.drexel.edu/~jbird/galflex/index.html}} 
module, which ray-trace second-order weak gravitational lensing 
effects (flexion), along with shear and convergence, on simulated 
source images in real space. While a full exploration of the 
11-dimensional parameter space is possible through combined 
downhill and sampling techniques (e.g. simulated annealing, etc.), 
for this work we take advantage of initial galaxy parameter 
estimates to aid global minimization.

\subsubsection{Goodness of Fit \& Parameter Uncertainties}
\label{sec:goodness_of_fit}

Already noted in Section \ref{sec:data_reduction}, the source-selection 
process is a critical part of flexion analysis. In a lensing reconstruction 
of Abell 1689 using ACS data, \citet{Leonard2007} combines strong 
lensing, shear, and flexion analysis to confirm substructure in the form 
of a second mass peak apart from the central cluster. However, the 
central mass peak is not recovered in their flexion analysis despite 
having a large mean number density of flexion sources 
(\(\bar{n}_g \sim 75\)/arcmin\(^2\).) A later flexion-only analysis of 
A1689 by \citet{Cain2011} is able to infer mass structure consistent 
with previous measurements with a mean flexion source density of 
\(\bar{n}_g \sim 26\)/arcmin\(^2\). \citet{Okura2008} finds that they 
can recover significant structure in the core of A1689 with only 9 
galaxies using Subaru data, through a meticulous source-selection 
procedure resulting in a final source density of 
\(\bar{n}_g \sim 7.75\)/arcmin\(^2\). The lack of a significantly stronger 
signal in the \citet{Leonard2007} analysis despite the larger source 
density indicates the necessity of selecting which galaxy sources 
effectively ``make the cut".

\begin{figure}
\begin{center}
\includegraphics[width=\imgwidth]{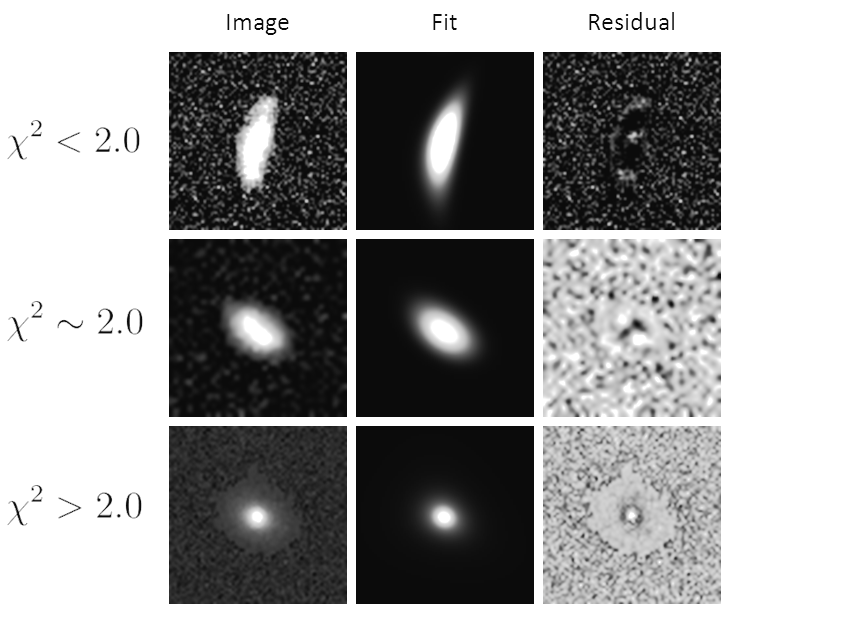}
\end{center}
\caption{Best-fit \Sersic{} galaxy models in the Abell 2744 Hubble Space Telescope F606w band. 
Each row shows a typical reduced \(\chi^2\) value used in part to group objects 
by a goodness-of-fit statistic. As systematic uncertainties in inherent model suitability, 
best-fit model error analyses, and model-parameter-error-size correlation 
are not yet well-understood, this statistic is not the defining accepted/rejected source
 criterion, but rather a weighted component of an extended source-selection pipeline.}
\label{fig:AIM_fit}
\end{figure}

Our final flexion source catalog is constructed by excluding the galaxies which 
have either poor or unphysical best-fit parameters. A commonly-used metric 
for determining a ``good" fit is the reduced chi-squared statistic 
\(\chi^2_{\mathrm{red}}\), defined as the weighted sum of squared 
deviations (the \(\chi^2\)) divided by the number of independent degrees 
of freedom of the system. Although we use \(\chi^2\) as the optimized 
statistic in AIM, we do not use the \(\chi^2_{\mathrm{red}}\) as the main 
selector for cutting our post-fit sources as in \citet{Cain2011} -- highly-correlated 
pixel data combined with a highly-nonlinear model render the actual 
meaning of the \(\chi^2_{\mathrm{red}}\) uncertain. We point the reader to 
\citet{Andrae2010} for further details. Nevertheless, this statistic is useful to 
broadly separate groups, as large values do indicate a likely poor fit. 
Figure \ref{fig:AIM_fit} shows an example of accepted, uncertain, 
and rejected galaxy source fits and their residuals, separated by 
\(\chi^2_{\mathrm{red}}\) for convenience.

Instead of the reduced chi squared statistic, we employ an impartial stamp inspection GUI as described in Section \ref{sec:stampcheck}, in combination with the following parameter cuts:
\begin{enumerate}
\item We reject fits where the reduced flexions \(\{|\Psi_1|, |\Psi_3|\} \ge 1.0\) arcsec\(^{-1}\). These large values indicate either a poor or unphysical fit, or that the source is in a region where weak lensing assumptions are not valid. \\
\item We also impose the constraint that \(\{\sigma_{\Psi_1},\sigma_{\Psi_3}\} > 0.001\) arcsec\(^{-1}\) to prevent overfitting the associated  parameters \(\{\Psi_1, \Psi_3\}\), as in \citet{Cain2011}. \\
\item We reject object fits with axis ratio \(q = 1.0\) in the same vein, as a best-fit which is exactly symmetric is likely to either be a star or overfitted. \\
\item Finally, for practical convenience we also eliminate objects whose best-fit \(\chi_{\mathrm{red}}^2\) is large (\(\gtrsim 5\)), in most cases representing either an improperly-deblended or foreground galaxy.
\end{enumerate}

\subsubsection{Interactive Limited Stamp Inspection}
\label{sec:stampcheck}

While the \texttt{Source Extractor} utility turns the nontrivial procedure of source identification, deblending, and shape estimation into something more routine for most purposes, the various strategy and parameter combinations - particularly for crowded fields with diverse redshift populations - can still prove a need for artful navigation. Parameters must be selected so as to identify both the small and faint sources, as well as the larger, brighter, and closer ones; as described in Section \ref{sec:data_reduction}, a common technique in weak lensing studies is to perform a ``hot-cold" routine. However, deblending issues still can persist throughout the pipeline. These issues can entail multiple galaxies masquerading as one, or vice versa, especially with extended galaxies containing luminous star-forming regions. 

Therefore, an additional layer of inspection can benefit a signal dependent on an accurate measure of a single galaxy's shape. While performing this kind of source inspection in a shear analysis would be time-prohibitive, this is not necessarily true for flexion analysis. As shown in \citet{Flexion/Viola}, accurate flexion measurements are particularly dependent on high signal-to-noise sources, and so by selecting objects according to suitable criteria the number of usable galaxies for flexion measurement is narrowed down to a much smaller and more feasible subset.

\begin{figure}
\begin{center}
\includegraphics[width=3.0in]{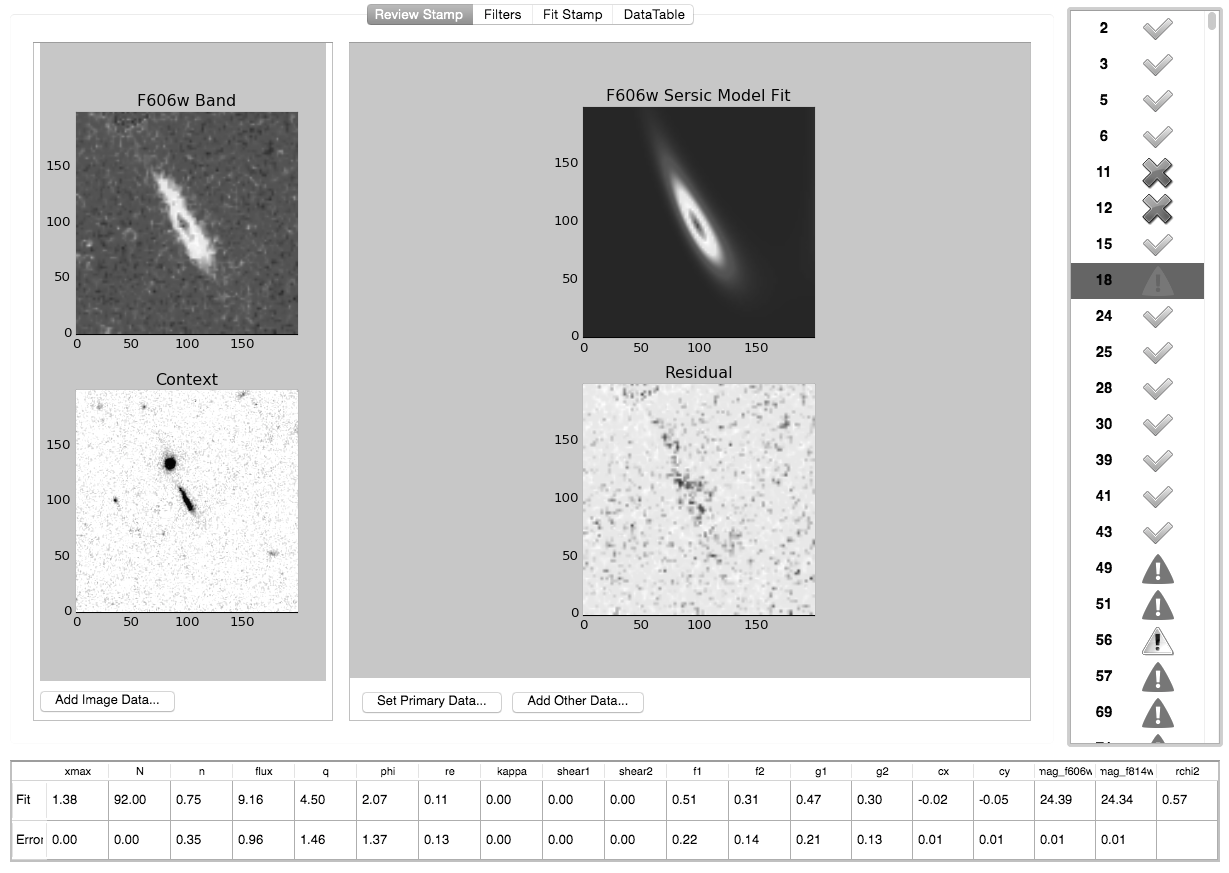}
\end{center}
\caption{Inspecting our image and best-fit stamp and fit parameters is critical for an analyses that can depend on just a few handful of strong signal images. We use a custom GUI to streamline the process of accepting/rejecting/flagging these sources for use in the mass reconstruction. }
\label{fig:stampcheck}
\end{figure}

Our stamp/fit inspection routine is implemented through a GUI as part of our flexion
pipeline. Built with the Python bindings for Qt (PyQt4), we compile the base code 
and all relevant modules into an executable to create a standalone application.
Seen in Figure \ref{fig:stampcheck}, the GUI can be used to quickly and 
efficiently scroll through input stamp datasets for acceptance/rejection. 
We show a slightly extended region around the selected stamp, in order to 
inspect its immediate neighbors for blending issues or galaxy fragmentation. 
We purposefully restrict the field of view in an attempt to minimize any observational bias. 
We also inspect the best-fit parameters to check for problematic behaviors, including having 
values on the established parameter bounds or exactly null, or having particular combinations 
of values that clearly indicate a poor fit (e.g. a best-fit half-light-radius larger than the stamp itself.)

\subsection{Mass Estimators}
\label{sec:reconstruction}

A major component of flexion analysis is the nontrivial process of incorporating flexion measurements into an overall mass reconstruction. Ideally, both shear and flexion measurements could be used to constrain the mass profile more than either could alone.  As we've noted, the shear probes large scales more effectively, while flexion is particularly suited to investigate smaller-scale structure. 

However, the flexion field alone can give valuable information about both small-scale structure in a lensing cluster and the overall mass profile. To do so, we employ two independent techniques to get a measure of a mass structure based solely on positional flexion measurements: a direct calculation of the convergence field, stemming directly from the fact that flexion is the gradient of the convergence; and the construction of a mass-signal map through the aperture mass technique \citep{Schneider1996}, which relates the weighted integral of flexion measurements within an aperture to the respective weighted integral of the convergence within that aperture \citep{Leonard2011,Cain2011}. Each technique offers a different interpretation of the flexion measurements, and the utilization of both can provide a broader image of any substructure present within the galaxy cluster.

\subsubsection{Direct flexion-convergence convolution}
Using the various directional derivative relations between lensing 
distortions and a gravitational potential, we are able to relate 
(up to the mass-sheet degeneracy) the convergence field 
\(\kappa\) with the flexion fields \(\F, \G\) through the convolution:
\begin{equation}\label{eqn:flexion_convolution}
\kappa(\bm{\theta}) = {\int_\mathbb{R}^2} d^2 \theta' \, \mathcal{D}_M (\bm{\theta} - \bm{\theta}') \, \F(\bm{\theta}')
\end{equation}
where \(\mathcal{M}\) is the \(\F\) or \(\G\) flexion field in the lens 
plane, and the respective convolution kernals are given by
\begin{align}
\mathcal{D}_{\F} \quad&=\quad \frac{\theta_1 + i\theta_2}{2\pi|\theta|^2} ,\\  
\mathcal{D}_{\G} \quad&=\quad \frac{(\theta_1 + i\theta_2)^3}{2\pi|\theta|^4}. 
\end{align}
as outlined in \citet{Leonard2011}. This can be applied to the case 
where all sources are assumed to be at the same redshift, or where 
the sources are distributed in redshift and the lensing fields are 
interpreted as those for sources at infinite redshift. In this work 
we use the latter definition, as the majority of our selected sources 
lack spectroscopic redshifts. 

While this convolution relation offers a direct linear inversion map 
with a straightforward interpretation, it suffers from the same 
difficulties as the shear convolution, namely that:
\begin{enumerate}
\item The integral in Eqn. \ref{eqn:flexion_convolution} extends over \(\mathbb{R}^2\) while usable flexion data is constrained to a few square arcminutes. \\
\item Smoothing over the lensing fields is necessary to avoid infinite noise from discrete sampling, which is a particular concern as the strength of the flexion signal falls off much more quickly than shear further from mass structure. \\
\item The cluster regions where flexion signal is expected to be strong are usually those where the convergence cannot be assumed to be small, and thus the observables are better approximated by the reduced flexion \(\Psi_1\) and \(\Psi_3\).
\end{enumerate}

To offer an alternative and complementary method to the direct convergence inversion, we also utilize our implementation of the aperture mass technique.

\subsubsection{Aperture Mass Statistic}

The aperture mass statistic is able to relate measured flexion to the underlying lens convergence through more general weighting kernals than the direct flexion-convergence convolution. Apertures of radius \emph{R} are laid down in a grid-like pattern over the field. Within each aperture the convergence is convolved with a filter function \(w(r)\). This convolution is then related to the measured flexion convolved with an appropriate filter function \(Q(r)\). The role of the aperture mass technique is to evaluate mass structure detections. By randomly rotating the flexion vectors in the image and running the reconstruction multiple times, it is possible to create a signal-to-noise map of detected mass signal, analogous to the flux signal-to-noise ratio (S/N) of a galaxy, for example.

The aperture statistic \(\fmap\){} is defined by relating the spin-1 flexion and convergence as follows:
\begin{align}
\fmap{}(\theta_0;R) &= \int_{|\theta|\leq R} d^2\, \theta \; \kappa(\theta+\theta_0)w(|\theta|) \\
\fmap{}(\theta_0;R) &= \int_{|\theta|\leq R} d^2\, \theta \; \mathcal{F}_E(\theta;\theta_0)Q(|\theta|)
\end{align}
where \(\mathcal{F}_E\) is the component of the first flexion oriented 
towards the center of the aperture. The weight and filter functions 
are free to be selected, while obeying the relations
\begin{equation}
Q(x) = -\frac{1}{x} \int_0^x w(x') x' dx'  \quad  w(x) = -\frac{1}{x} Q(x) - \frac{dQ}{dx} .
\end{equation}
The weight function is constrained to go to zero smoothly at the 
aperture boundary and have a mean value of zero,
\begin{equation}
\int_0^R w(x') x' dx' = 0.
\end{equation}
In this work we use a family of polynomial weighting functions as described 
and used in the literature \citep{Schneider1998,Leonard2011,Cain2011}. 
These filter functions are chosen to test the persistence of mass estimates 
across a variety of scales. 
The convergence weighting function is defined as
\begin{equation}
w(x) = A_l \frac{(2 + l)^2}{\pi} \left(1 - \frac{x^2}{R^2}\right)^l \left(\frac{1}{2+l} - \frac{x^2}{R^2}\right)
\end{equation}
along with the flexion kernal
\begin{equation}
Q(x) = -A_l \frac{2 + l}{2\pi} x \left(1 - \frac{x^2}{R^2}\right)^{1+l}
\end{equation}
and normalization
\begin{equation}
A_l = \frac{4}{\sqrt{\pi}}\frac{\Gamma(\frac{7}{2} + l)}{\Gamma(3 + l)}
\end{equation}
A higher polynomial index makes the kernel more sensitive to 
smaller scale structure within the aperture, at the expense of 
having larger noise fluctuations. Lower indices smooth over a 
larger area, reducing noise at the cost of resolution. 
The aperture radius has a similar effect: as the radius increases 
the flexion signal associated with the center of the aperture falls 
off quickly, smoothing over smaller scale structures.


\section{Summary and Results}
\label{sec:results}

\subsection{Source Selection Results}
\label{sec:source_cuts}

Our extraction and pre-measurement selection routine returned 1344 out 
of the 4249 detected galaxies in the F814w-observed 3.5' x 3.5' field of Abell 2744.
The after-measurement selection cuts give us our final catalog of 969 
sources, corresponding to 82.1 sources arcmin\(^{-2}\).

In the cluster's associated parallel field our pre-selection routine 
returned 1001 out of 4142 detected galaxies. Post-measurement 
cuts reduce our final catalog to 923 usable galaxies, corresponding to 
78.2 sources arcmin\(^{-2}\). 
As the observational details of the parallel field are identical to those 
of the cluster (e.g. a field of the exact same size, with the same total 
integration time in each band, in the same number orbits), a source 
density comparable to the cluster field conforms with the expectation 
that our cluster selection routine is able to exclude contaminating 
cluster members.

\subsection{Mass Reconstruction Results}

\begin{figure*}
\begin{center}
\includegraphics[width=7in]{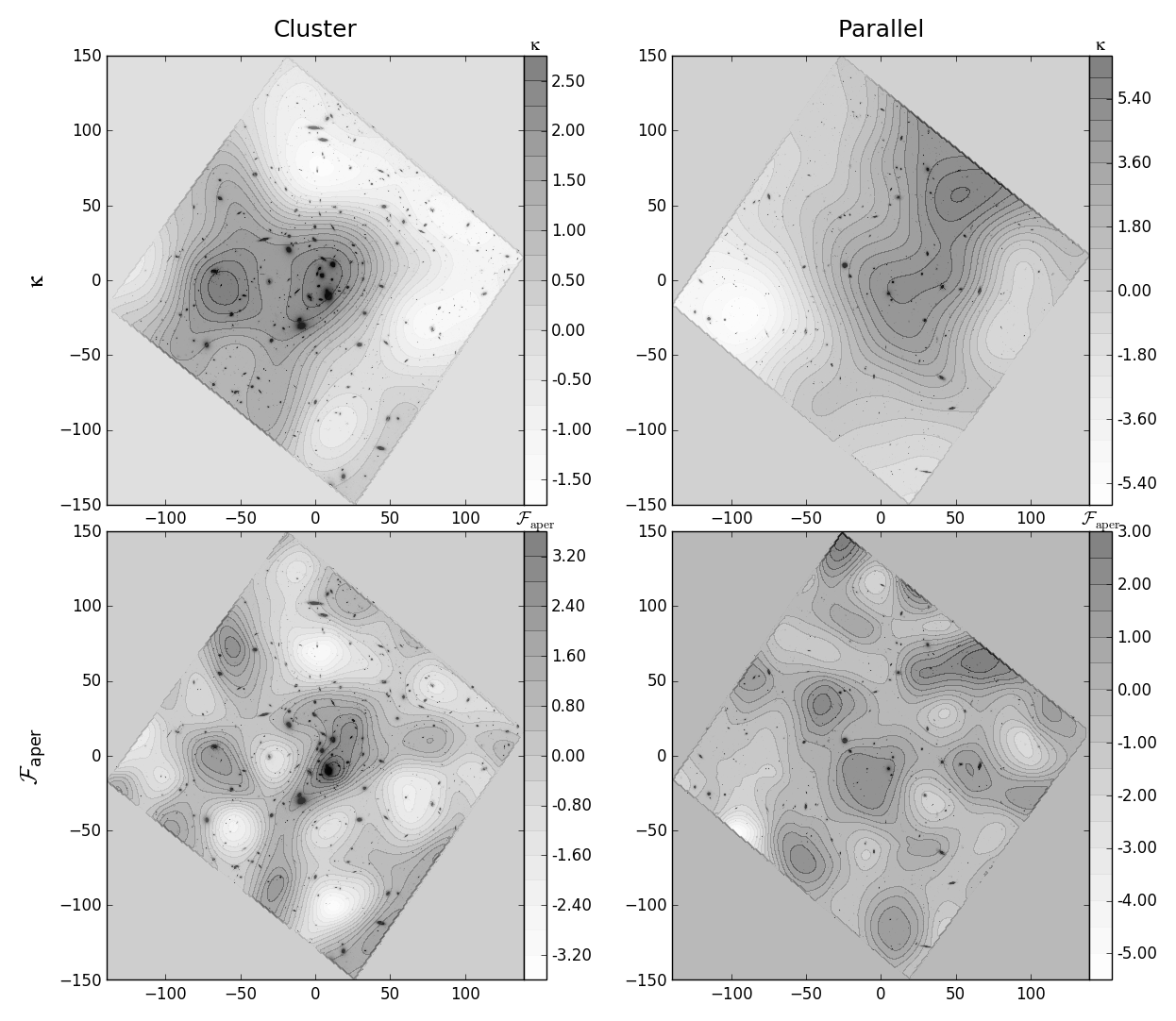}
\end{center}
\caption{Independent \(\F\)-flexion data is used to create cluster-scale mass field 
reconstructions of Abell 2744 and noise in its associated parallel field. The top left 
figure shows the cluster convergence (up to a mass-sheet degeneracy scaling) 
using a 15" smoothing kernal, while the top right shows a parallel field noise 
reconstruction under the same parameters. The bottom left and right figures 
are the flexion aperture mass signal-to-noise reconstructions for the cluster 
and parallel fields, respectively, at aperture radius 60" and index \(\ell=5\). 
The axes are in units of arcseconds.}
\label{fig:flexion_panel}
\end{figure*}

Figure \ref{fig:flexion_panel} presents one of our cluster-scale mass 
reconstructions of Abell 2744 as well as the parallel field noise 
reconstruction, using \(\F\)-flexion data exclusively. 
The top left graph shows the direct flexion-convergence cluster 
reconstruction with flexion smoothing scale \(\sigma = 15"\). 
There are two distinct peaks at this resolution -- a large central peak 
that follows the elliptical luminosity structure of the central cluster 
galaxies, declining smoothly towards the eastern edges, and 
a lesser second peak centered just south of a foreground galaxy to the west,
offset 1.43' (\(273h^{-1}\) kpc) from the cluster core. 

To estimate the mass of these structures the mass-sheet degeneracy 
must be broken, and we do so using two independent methods.
The most direct way to constrain the \(\kappa\) reconstruction is 
by scaling the data to go to a known value at the field boundaries, under
the transformation \(\kappa' = \lambda \kappa + (1-\lambda\)) in the nonlinear 
weak lensing regime.
Based on the radial behavior of typical lensing models of clusters of similar properties 
(e.g. mass, Einstein radius), we constrain the smooth 
eastern edge of the data to \(\kappa = 0.1\) and scale the convergence 
map accordingly. 
For the primary peak we find a mass of 
\(1.93\times10^{14}\,h^{-1}\,M_{\odot}\) within a 45" 
aperture, corresponding to 145\(h^{-1}\) kpc 
(200 kpc assuming a h=0.7 cosmology), and 
for the western substructure we find a mass of 
\(7.12\times10^{13}\,h^{-1}\,M_{\odot}\) within a 25" 
aperture, corresponding to 80\(h^{-1}\) kpc  
(110 kpc assuming a h=0.7 cosmology).
While assuming a convergence value at the field boundary 
could significantly affect an estimate of the total integrated 
cluster mass \citep[e.g.][]{Bradac/Lombardi2004}, the effect 
on our measured peak mass within an aperture is 
limited -- for example, increasing or reducing the assumed 
boundary convergence by \(\kappa \pm 0.1\) results in a contained
mass difference of \(\mp 0.04\times10^{14}\,h^{-1}\,M_{\odot}\) 
for our primary structure, or 2\% of our calculated mass, 
with similar results (3\%) for the secondary peak.

The mass-sheet degeneracy is also able to be constrained by fitting a lens model
to the data. We fit a softened isothermal sphere profile
to each of the identified peak locations, allowing the center coordinates to 
marginally vary while the model velocity dispersion is optimized against our flexion data.
The large central peak converges to a best-fit with velocity dispersion 
\(\sigma_v = 1630\) km/s, while the offset substructure peak converges to a 
model with  \(\sigma_v = 766\) km/s.

Comparing with other estimates of Abell 2744, in a strong lensing analysis
\citet{Jauzac2014} found a contained mass of 
\(M(<200 kpc) = (2.162 \pm 0.005) \times 10^{14} M_{\odot}\) within a 200 kpc 
aperture, which is similar to our estimate of the central potential.
Although our relatively simple and straightforward mass calculation does not lie within the 
error bars of the precise strong lensing estimate, \citet{Jauzac2015_2} 
stresses that the precision of the mass models 
from gravitational-lensing studies depends strongly on the mass modeling 
technique, and mass estimates from different groups using different strong 
lensing algorithms will find different results.

The parallel field convergence noise reconstruction under the same 
parameters is shown in the top right graph of Figure \ref{fig:flexion_panel}. 
Compared to the cluster mass reconstruction, the derived magnitudes are 
larger but the noise peaks are much broader and do not correlate with any 
luminous matter. In addition, the magnitude of the highest value is 
approximately the same as that of the lowest, unlike our derived 
cluster convergence map. As we expect the parallel field to have 
\(\langle\kappa\rangle \sim 0\), the observed \(|\kappa|\) parity around 0.0 
is an indicator of the magnitude of systematic noise in an unconstrained 
flexion-convergence reconstruction.

While convergence-flexion relation allows us to calculate 
physical properties of the cluster or peaks inside a given aperture 
(provided the convergence is normalized to the mass-sheet degeneracy), 
the aperture mass method gives a more qualitative result of where 
mass peaks lie. The lower left and lower right graphs show the cluster/parallel aperture 
mass/noise reconstructions using aperture R = 60" and polynomial 
index \(\ell = 5\).
Mass signal-to-noise contours in the cluster indicate a strong central peak 
aligned and centered on the BCG and extending along the cluster's luminosity 
semi-major axis. 
A corresponding lesser peak is identified to the west along with another to the northwest. 
The signal tapers to zero near the edge of the data, with the 
exception of a few spurious structures along the edge.  
The parallel field map does not appear to have any significant peaks, beyond 
apparent edge effects (specifically at the north and northeast edges), nor do 
the minor peaks detected within the field (S/N \(\sim 1\)) correlate with any luminous structure present.

We also compare the scales of structural predictions between flexion and shear by 
producing convergence reconstructions through direct shear inversion. Although our
treatment of the PSF (using a simple Gaussian) would likely introduce ellipticity 
bias in an analysis on much wider scales, we expect a much stronger shear
signal in the inner 3.5' x 3.5' cluster field. 

\begin{figure*}
\begin{center}
\includegraphics[width=7in]{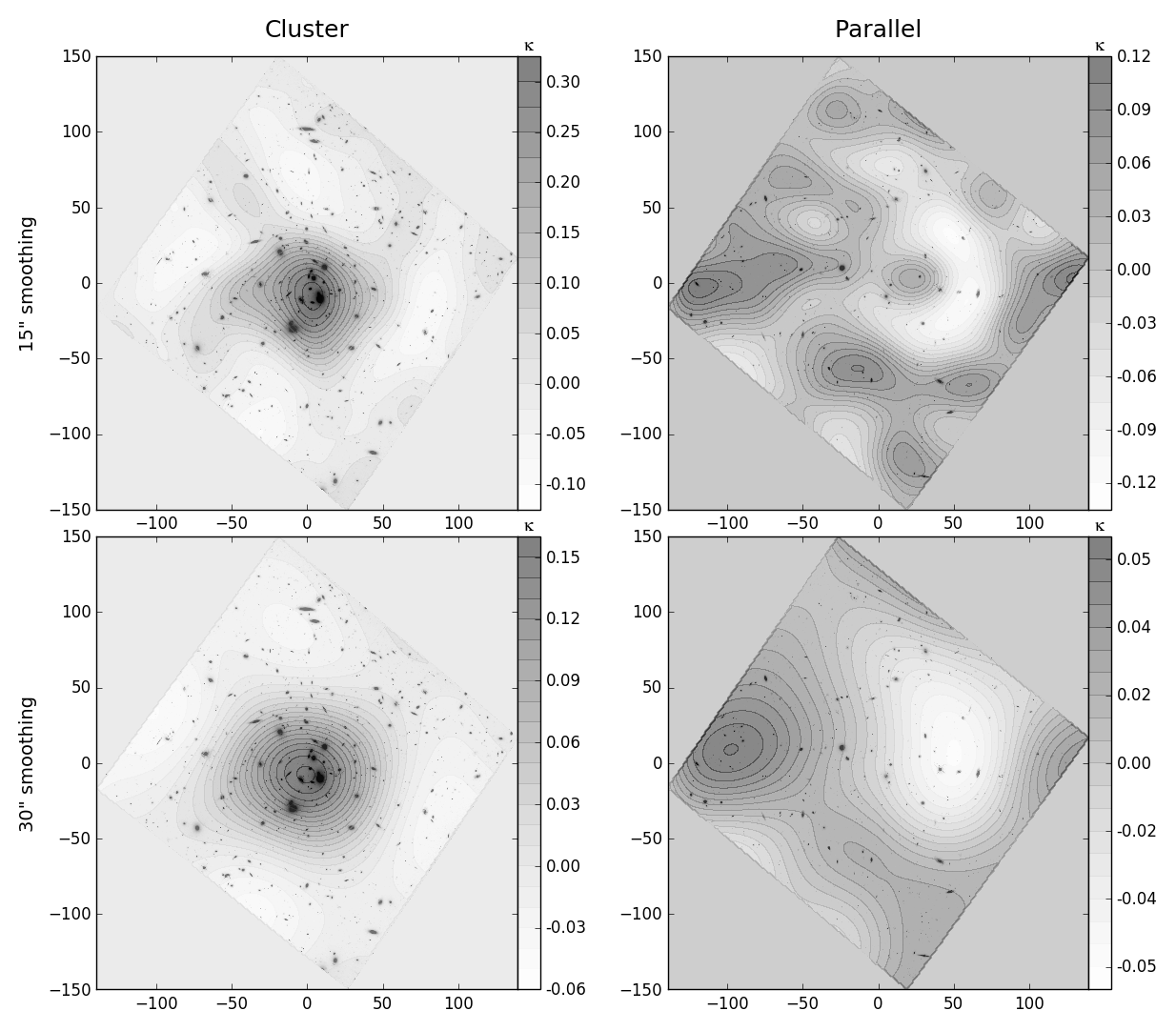}
\end{center}
\caption{Independent AIM-derived shear data is used in a cluster-scale mass 
field reconstruction of Abell 2744 and a noise reconstruction in its associated 
parallel field (both up to a mass-sheet degeneracy scaling). The top left 
and right figures show the convergence field resulting from a 15" smoothing 
kernal applied to the cluster and parallel fields, respectively, while the bottom 
left and right figures display the results while using a 30" kernal instead. 
The axes are in units of arcseconds.}
\label{fig:shear_panel}
\end{figure*}

Figure \ref{fig:shear_panel} shows our
convergence reconstructions using just the averaged galaxy ellipticities measured 
through AIM fitting, again up to a mass-sheet degeneracy. Following a similar format, 
the top two graphs show the cluster and parallel field \(\kappa\) maps using a 15" 
shear smoothing kernal, while the bottom graphs utilize a 30" shear kernal. 
The clear broad central structure of the cluster peak at each scale is 
apparent, and both estimate the cluster centroid to be very 
close to what the flexion maps predict. 
However, both shear smoothing scales do not show any strong indications of 
local substructure.
In the 15" map the convergence extends slightly towards and hints at the western and 
northwestern structures highlighted by our flexion analysis, but does not 
reproduce the flexion map's well-defined structural peaks, while the 
30" reconstruction shows no indication of possible local structure at all.

To test the PSF assumptions and reconstruction method, we produced noise 
reconstructions of the shear in the parallel field in the same manner as 
the flexion procedure. Neither parallel field shear smoothing scale shows 
any appreciable structural significance, particularly when 
compared with the strong central profile of the cluster field, and 
the overall noise signal is centered at 0.0 similar to the flexion parallel field 
reconstructions. 

While the gravitational shear field can be used effectively to determine 
the overall mass structure of galaxy clusters, 
its extended nonlocal effects as well as its inherent ellipticity-degeneracy 
limit its use to broad mass distributions, and thus it 
is not a viable candidate for higher resolution substructure detections.
Additionally, while strong lensing can and has led to precise characterizations 
of inner cluster cores, multiply-imaged source systems are not guaranteed to 
be located near substructure and quickly become sparse outside the dense core. 
As an intermediary signal spanning the strong and weak lensing regimes, 
gravitational flexion signal has the ability to effectively probe significant cluster 
substructure on scales and at angular extents which cannot be practically 
detected through other means. 
As inherent flexion noise and systematic bias become more well-understood, 
flexion signal has the potential to be a key component in both exploring 
the behavior of galaxy cluster formation and evolution as well as 
understanding the nature of dark matter structural dynamics.

In this work we have used an Analytic Image Modeling implementation (AIM) 
to measure flexion signal in the Abell 2744 galaxy cluster and inherent flexion 
in Abell 2744's associated parallel field. We show the efficacy of using flexion 
alone as an indicator of structure, exploring a much deeper view into the 
inner core of the cluster than shear would allow, and investigate the role 
of different mass estimators in both the cluster and parallel field. 
We identify and obtain mass estimates for both the central core of the cluster 
and a detected substructure offset 1.43' to the west of the core.  
Finally, we demonstrate that we are able to make simultaneous 
measurements of the shear field while measuring flexion through 
the AIM technique, and reconstruct the broader cluster mass 
structure while finding no such signal in the parallel field.

\bigskip

Based on observations obtained with the NASA/ESA Hubble Space Telescope, 
retrieved from the Mikulski Archive for Space Telescopes (MAST) at the Space 
Telescope Science Institute (STScI). STScI is operated by the Association of 
Universities for Research in Astronomy, Inc. under NASA contract NAS 5-26555. 

\bibliographystyle{mn2e_mod2}
\bibliography{birdgoldberg_sources}

\newpage

\bigskip
\newpage

\end{document}